\RequirePackage[hyphens]{url} 
\documentclass[twocolumn,tighten,times]{aastex63}
\usepackage{amssymb}
\usepackage{amsmath}
\usepackage{amsfonts}
\usepackage{comment}
\usepackage{graphicx}
\usepackage{txfonts}
\usepackage{enumitem}
\usepackage{xspace}
\usepackage{multirow, makecell}

\def\be{\begin{equation}}
\def\ee{\end{equation}}
\def\ba{\begin{eqnarray}}
\def\ea{\end{eqnarray}}

\newcommand{\DM}{\ensuremath{\mathrm{DM}}}
\newcommand{\DMhat}{\ensuremath{\widehat{\mathrm{DM}}}}

\newcommand{\Dtd}{\Delta t_{\mathrm{d}}}
\newcommand{\Dnud}{\Delta \nu_{\mathrm{d}}}
\newcommand{\taud}{\tau_{\mathrm{d}}}
\newcommand{\niss}{n_{\mathrm{ISS}}}

\newcommand{\snr}{S/N}

\newcommand{\J}{J0437$-$4715\xspace}
\newcommand{\PSR}{PSR~J0437$-$4715\xspace}

\newcommand{\Weff}{W_{\rm eff}}
\newcommand{\Uobs}{U_{\rm obs}}
\newcommand{\Nphi}{N_\phi}
\newcommand{\dt}{{\delta t}}
\newcommand{\fL}{f_{\rm L}}
\newcommand{\fH}{f_{\rm H}}

\newcommand{\W}{\ensuremath{\mathrm{W}}}
\newcommand{\C}{\ensuremath{\mathrm{C}}}

\newcommand{\SB}{{\mathrm{SB}}}

\newcommand{\A}{{\mathrm{A}}}
\newcommand{\R}{{\mathrm{R}}}

\newcommand{\T}{\mathcal{T}}

\shorttitle{Observations of \PSR with the IAR}
\shortauthors{Lam \& Hazboun}

\begin{document} 
\title{Precision Timing of \PSR with the IAR Observatory and Implications for Low-Frequency Gravitational Wave Source Sensitivity}

\author[0000-0003-0721-651X]{M.\,T.\,Lam}
\affiliation{School of Physics and Astronomy, Rochester Institute of Technology, Rochester, NY 14623, USA; mlam@mail.rit.edu}
\affiliation{Laboratory for Multiwavelength Astrophysics, Rochester Institute of Technology, Rochester, NY 14623, USA}
\author[0000-0003-2742-3321]{J.\,S.\,Hazboun}
\affiliation{Physical Sciences Division, University of Washington Bothell, 18115 Campus Way NE, Bothell, WA 98011, USA}

\begin{abstract}

While observations of many high-precision radio pulsars of order $\lesssim1~\mu$s across the sky are needed for the detection and characterization of a stochastic background of low-frequency gravitational waves (GWs), sensitivity to single sources of GWs requires even higher timing precision. The Argentine Institute of Radio Astronomy (IAR; Instituto Argentino de Radioastronom\'{i}a) has begun observations of the brightest-known millisecond pulsar, J0437$-$4715. 
Even though the two antennas are smaller than other single-dish telescopes previously used for pulsar timing array (PTA) science, the IAR's capability to monitor this pulsar daily coupled with the pulsar's brightness allows for high-precision pulse arrival-time measurements. While upgrades of the facility are currently underway, we show that modest improvements beyond current plans will provide IAR with unparalleled sensitivity to this pulsar. The most stringent upper limits on single GW sources come from the North American Nanohertz Observatory for Gravitational Waves (NANOGrav). Observations of \PSR will provide a significant sensitivity increase in NANOGrav's ``blind spot'' in the sky where fewer pulsars are currently being observed. With state-of-the-art instrumentation installed, we estimate the array's sensitivity will improve by a factor of \replaced{$\approx$2-7}{$\approx$2-4} over 10 years for 20\% of the sky with the inclusion of this pulsar as compared to a static version of the PTA used in NANOGrav's most recent limits. More modest instrumentation result in factors of \replaced{$\approx$1.4-4}{$\approx$1.4-3}. We identify four other candidate pulsars as suitable for inclusion in PTA efforts. International PTA efforts will also benefit from inclusion of these data given the potential achievable sensitivity.

\end{abstract}


\section{Introduction}
\label{sec:intro}

The search for gravitational waves (GWs) using an array of recycled millisecond pulsars (MSPs) is a key science goal for the largest radio telescopes in the world. Pulsars
act as high-precision clocks and the passage of a GW along the line of sight is expected to cause slight variations in the arrival times of their observed emission that can be
measured using high-precision pulsar timing \citep{Sazhin1978,Detweiler1979,fb1990}.  Large telescopes are needed to measure high signal-to-noise ratio (S/N) pulses and estimate their arrival times to high precision \citep{Ransom+2019}. While current pulsar-timing efforts are dominated by large single-dish (60- to 300-m-class) telescopes, the use of larger, more sensitive telescopes and interferometers is becoming an increasingly important contribution to  pulsar timing array (PTA) science \citep{IPTADR2}.

The expected first detection of low-frequency ($\sim$nHz-
$\mu$Hz) GWs is from a stochastic background of unresolved supermassive black hole binaries \citep[e.g.,][]{Rosado+2015}. Detection and characterization of this background require observations of many pulsars to build pairwise quadrupolar correlations in the observed arrival-time perturbations. The best-timed MSPs are used in these searches, with timing precision $\lesssim$1~$\mu$s. Individual sources of GWs, such as from a single resolved supermassive black hole binary inspiral or merger, require MSPs with precision an order-of-magnitude or \replaced{better}{more} such that specific waveforms can be recovered \citep[e.g.,][]{blf2011}. \added{Red noise in the pulsar TOAs, e.g., such as from intrinsic spin noise or the GW background, will impact the waveform recovery but mostly at the lowest GW frequencies \citep[$\lesssim 10$~nHz;][]{Xin+2020}; we discuss this more later with respect to our analysis.}
Observing many well-timed pulsars
both increases the signal-to-noise ratio of a GW detection and future characterization, and improves upon the coverage of events
across the sky.

\subsection{The North American Nanohertz Observatory for Gravitational Waves}

The North American Nanohertz Observatory for Gravitational Waves (NANOGrav; \citealt{McLaughlin2013,Ransom+2019}) collaboration is working towards the detection and characterization of low-frequency GWs by means of monitoring an array of nearly 80 precisely-timed MSPs using three telescopes: the 305-m William E. Gordon Telescope of the Arecibo Observatory (Arecibo), the 100-m Robert C. Byrd Green Bank Telescope of the Green Bank Observatory (GBO), and the 27-element 25-m antennas of the National Radio Astronomy Observatory's (NRAO) Karl G. Jansky Very Large Array (VLA). NANOGrav has already placed upper limits on the amount of GWs in the Universe, which have provided strong astrophysical constraints on merging galaxies and cosmological models \citep{NG11GWB,NG11CW,NG11BWM,3C66B}. NANOGrav is poised to detect and begin characterization of the stochastic GW background from an
ensemble of unresolved supermassive black hole binaries (SMBHBs) within the next 3-5 years (\citealt{Rosado+2015,Taylor+2016,NG11GWB}; Arzoumanian et al. in prep). The first detection of a continuous-wave (CW) source of GWs from a single SMBHB is expected by
the late 2020s \citep{Rosado+2015,Mingarelli+2017,Kelley+2018,NG11CW}. In addition, NANOGrav is sensitive to the direct signature of SMBHB mergers, known as a ``burst with memory'' (BWM; \citealt{NG11BWM}).

\subsection{The Argentine Institute for Radio Astronomy}

The Argentine Institute of Radio Astronomy (IAR; Instituto Argentino de Radioastronom\'{i}a) has begun upgrades of two 30-meter antennas primarily for the purpose of pulsar observations \citep{IAR}. Located at latitude $\approx -37^\circ$,
the observatory covers declinations $\delta < -10^\circ$. The two antennas, A1 and A2, are 30-m equatorial-mount telescopes separated by 120 m. Both are capable of observations around 1.4~GHz, each with current specifications listed in Table~\ref{table:A1A2}.

While 30-meter antennas fall below the typical size of radio telescopes used in precision timing experiments \citep[e.g.,][]{IPTADR2}, the IAR Observatory has several advantages for high-precision timing, specifically in the area of GW detection and characterization. With access to the Southern sky below $\delta < -10^\circ$ and many galactic-disk pulsars, and the ability to track objects for nearly four hours, IAR can and has demonstrated the capability of high-cadence timing on a number of those pulsars, specifically of the bright millisecond pulsar \J \replaced{\citep{IAR}}{\citep{IAR,IAR2}}. Their preliminary observations suggest that future measurements will be a significant contributor to PTA sensitivity.

\begin{deluxetable}{l|cc}
\tablecolumns{3}
\tablecaption{Current Observing Paramters for the IAR Antennas}
\tablehead{
\colhead{Parameter} & \colhead{A1} & \colhead{A2}
}
\startdata
Maximum Tracking Time ($T_{\rm obs}$) & 3 hr 40 min & 3 hr 40 min\\
Circular Polarizations ($N_{\rm pol}$) & 1 & 2\\
Receiver Temperature ($T_{\rm rcvr}$) & 100~K & 110~K \\
Gain ($G$) & 11.9 Jy/K & 13.0 Jy/K\\
Frequency Range (MHz) & 1100--1510 & 1200--1600 \\
Bandwidth ($B$) & 112 MHz & 56 MHz
\enddata
\tablecomments{All values taken from \citet{IAR}. \added{\PSR can be observed for the maximum tracking time.}}
\label{table:A1A2}
\end{deluxetable}

\subsection{The Millisecond Pulsar \J}

Discovered in the Parkes Southern Pulsar Survey \citep{Johnston+1993}, \PSR is the brightest-known MSP and has the lowest dispersion measure (DM; the integrated line-of-sight electron density) of any MSP; it is the second-lowest of any pulsar \citep{PSRCAT}. \replaced{It has a close distance of $156.8 \pm 0.3$~pc measured from the derivative of the binary orbital period \citep{Reardon+2016}. As such, high S/N pulses can be precisely timed, making this pulsar an excellent addition to PTAs.}{Its bright pulses can be precisely timed with high S/N, allowing for this pulsar to be used in tests of general relativity \citep{vS+2001,Verbiest+2008} and pulsar emission physics \citep{Oslowski+2014}, as a source in the development of new high-precision timing calibration and methodology \citep{Liu+2011,vS2006,Oslowski+2013}, and as a sensitive component to pulsar timing arrays \citep{IPTADR2,Kerr+2020}.}  Some pulsar properties are listed in Table~\ref{table:J0437}.

In addition to the ability to measure high S/N pulses, given its low DM, we expect that unmodeled chromatic (i.e., radio-frequency-dependent) interstellar propagation effects are reduced for this pulsar \citep{cs2010}, thereby limiting one of the significant sources of timing noise found in many other pulsars. \added{PSR~\J, however, is not timed without complications.} \deleted{While} The pulsar has measurable achromatic noise consistent with rotational spin fluctuations \citep{sc2010,NG9EN}, \replaced{its high-precision timing nonetheless allows it to be used in tests for deterministic GW signals.}{quasi-periodic intensity and polarization variations are seen in sub-pulse components \citep{De+2016}, and a systematic variation in the pulse shape has also been observed \citep{Kerr+2020}. While these effects complicate the pulsar's use in tests of fundamental physics with high-precision timing, nonetheless its demonstrated stability suggests that it can be used in these tests \citep[e.g.,][]{Verbiest+2009}, including in searches for stochastic and deterministic GW signals.
}

In this paper, we demonstrate the \replaced{unique}{specific} capabilities of the IAR to observe \PSR and provide unparalleled sensitivity to the pulsar for the purpose of single-source CW detection and future characterization. In \S\ref{sec:sensitivity}, we provide the framework for our time-of-arrival (TOA) sensitivity estimates. We consider the current IAR observations of \PSR in \S\ref{sec:current_obs} and then extrapolate to future possible operating modes in \S\ref{sec:future_obs}, comparing the modes to observations of the pulsar by other international observatories as well. In \S\ref{sec:projections}, we demonstrate how the IAR will improve sensitivity to single-source CWs when the data are combined with that of NANOGrav. As the collaboration observing the most \deleted{number of} pulsars \citep{Ransom+2019} and with the most recent and stringent CW limits \citep{NG11CW,3C66B}, we will focus our analyses primarily on improving upon a {\it projected} NANOGrav data set with observations of \PSR, noting that upcoming analyses (e.g., with \citealt{NG12,NG12WB}, which already include more time baseline and more pulsars) imply greater sensitivity in the future. We describe other possible pulsar targets in \S\ref{sec:other_MSPs} and conclude in \S\ref{sec:conclusion}.

\section{The Sensitivity toward Continuous Waves}
\label{sec:sensitivity}

In this section, we describe the mathematical framework for determining the precision of an individual pulsar and then the sensitivity toward CW sources from the sum of correlated GW signals between pulsar pairs. As the components of this framework have been discussed significantly throughout the literature, we will lay out each piece but \deleted{only} briefly, and will point the reader to other references with more detailed discussion.

\subsection{Single-Pulsar Timing Sensitivity}

\begin{deluxetable}{l|c}
\tablecolumns{2}
\tablecaption{Parameters for \PSR}
\tablehead{
\colhead{Parameter} & \colhead{Value}
}
\label{table:J0437}
\startdata
Spin Period$^{\rm a}$ ($P$) & 5.76 ms\\
Dispersion Measure$^{\rm a}$ (DM) & 2.64~pc~cm$^{-3}$\\
Effective Width$^{\rm b}$ ($\Weff$) & 667.6 $\mu$s\\
Flux Density at 400~MHz$^{\rm a}$ ($S_{400}$, ) & 550.0 mJy\\
Flux Density at 1000~MHz$^{\rm c}$ ($S_{400}$)  & 223 mJy\\
Flux Density at 1400~MHz$^{\rm a}$ ($S_{1400}$) & 160.0 mJy\\
Spectral Index$^{\rm c}$ ($\alpha$) & $-$0.99 \\
Jitter rms for 1 hour at 730~MHz ($\sigma_{\rm J,730}$)  & $61 \pm 9$ ns\\
Jitter rms for 1 hour at 1400~MHz ($\sigma_{\rm J,1400}$)  & $48.0 \pm 0.6$ ns\\
Jitter rms for 1 hour at 3100~MHz ($\sigma_{\rm J,3100}$)  & $41 \pm 2$ ns\\
\hline
Phase Bins ($\Nphi$, assumed) & 512\\
Median $\sigma_{\rm S/N}$ for Antenna 1 ($\sigma_{\rm S/N, A1}$) & 200 ns \\
Median $\sigma_{\rm S/N}$ for Antenna 2 ($\sigma_{\rm S/N, A2}$) & 330 ns \\
Median S/N for Antenna 1$^{\rm c}$ ($\tilde{S}_{\rm A1}$) & 148.8 \\
Median S/N for Antenna 2$^{\rm c}$ ($\tilde{S}_{\rm A2}$) & 90.5 \\
\enddata
\tablecomments{$^{\rm a}$Values taken from \textsc{psrcat} \citep{PSRCAT}. $^{\rm b}$Effective width estimated from profile in \citet{Kerr+2020}. $^{\rm c}$Derived parameters.}
\end{deluxetable}

We use the framework of \citet{optimalfreq} to estimate the noise contributions to pulsar TOA uncertainties, or equivalently how precisely we can measure the TOAs. \citet{IAR} show \added{the first set of} timing residuals (the observed TOAs minus a model for the expected arrival times) for \PSR\ \added{with the IAR} and demonstrate excess noise beyond the uncertainties derived from the common practice of matching a pulse template to the observed pulsar profile to estimate their TOAs \citep{Taylor1992}.

\subsubsection{Short-timescale Noise}
\label{sec:shorttimenoise}

\citet{NG9WN} describe a model for white noise in timing residuals on short timescales, $\lesssim 1$~hr, i.e., the time of typical observations. In that analysis, they assume perfect polarization calibration and radio-frequency-interference (RFI) removal; the former is currently impossible in the case of A1 with only one measured polarization channel. However, assuming both are true, the three white-noise components described are the template-fitting, jitter, and scintillation noise terms. 

Template-fitting errors rely on the matched-filtering assumption that the observed pulse profiles are an exact match to a template shape with some additive noise. These are the {\it minimum} possible errors for TOA estimates, and are given by
\be 
\sigma_{\rm S/N}(S) = \frac{\Weff}{S\sqrt{\Nphi}},
\label{eq:TFerror}
\ee
where $\Weff$ is the effective width of the pulse, $\Nphi$ is the number of samples (bins) across the profiles, and $S$ is the signal-to-noise ratio (written this way in equations for clarity), taken to be the peak to off-pulse rms. The effective width is related to the spin period of the pulsar $P$ and the template profile normalized to unit height, $U(\phi)$, by
\be 
\Weff = \frac{P}{\Nphi^{1/2} \left[\sum_{i=1}^{\Nphi-1} \left[\Uobs(\phi_i) - \Uobs(\phi_{i-1})\right]^2\right]^{1/2}}.
\label{eq:Weff}
\ee
Pulse profiles are frequency-dependent and so this error will apply to each profile measured at a given frequency. However, given the low-bandwidths currently available at the IAR Observatory, it is sufficient given the S/N-regime of the observations to average the data across not only time but frequency as well to obtain a single TOA per epoch per antenna. Similarly, many other frequency-dependent variations within the band will be small and we will ignore those here.

While $\Weff$ and $\Nphi$ are constants, the pulse S/N will vary on short timescales due to diffractive scintillation, with a known probability density function (PDF) written as \citep{cc1997,NG9WN} 
\be
f_S(S \vert S_0, \niss) = \frac{(S\niss/S_0)^{\niss}}{S\Gamma(\niss)} e^{-S\niss/S_0}\Theta(S),
\label{eq:sn_pdf}
\ee
where $S_0$ is the mean S/N, $\Gamma$ is the Gamma function, $\Theta$ is the Heaviside step function, and $\niss$ is the number of scintles in the observation of length $T$ and bandwidth $B$, given by
\be
\niss \approx \left(1+\eta_t \frac{T}{\Dtd}\right)\left(1+\eta_\nu \frac{B}{\Dnud}\right).
\label{eq:N_DISS}
\ee
The scintillation parameters $\Dtd$ and $\Dnud$ describe the characteristic timescale and bandwidth of intensity maxima, or scintles, in a dynamic spectrum. The $\eta_t$ and $\eta_\nu$ parameters are the scintle filling factors $\approx 0.2$ \citep{cs2010,Levin+2016}.

\citet{Gwinn+2006} found two diffractive scintillation scales for \PSR made at an observing frequency of 328~MHz. The first has $\Dtd = 1000$~s and $\Dnud = 16$~MHz while the second has $\Dtd = 90$~s and $\Dnud = 0.5$~MHz. For a Kolmogorov medium, we can scale these quantities as $\Dtd \propto \nu^{6/5}$ and $\Dnud \propto \nu^{22/5}$ \citep{cr1998, optimalfreq}. Therefore, when scaled to an observing frequency of 1.4~GHz, we have $\Dtd = 5700$~s and $\Dnud = 9.5$~GHz for the first scale, and $\Dtd = 510$~s and $\Dnud = 300$~MHz for the second scale. \citet{Keith+2013} measured the scintillation parameters at 1.5~GHz for \PSR; when scaled to 1.4~GHz, they found $\Dtd = 2290$~s and $\Dnud = 740$~MHz, in between the two scales observed by \citet{Gwinn+2006}.

Using these measurements of the scintillation parameters, we will estimate $\niss$ and thus the predicted impact on the PDF of $S$. Given the current small bandwidths of the receiver, since $B < \Dnud$ in all cases, the second of the two components in Eq.~\ref{eq:N_DISS} will be $\approx 1$ and so we have $\niss \approx 1 + \eta_t T/\Dtd$. For $T = 13200$~s, this quantity will range between 1.5 and 5.2. The higher $\niss$ is, the more that the PDF will tend towards the mean S/N value $S_0$, whereas when $\niss$ is close to 1, the distribution becomes exponential. Therefore, this value will heavily dictate the timing uncertainties achievable, which we will briefly discuss when applying to the real data.

The two other short-timescale contributions to the white noise discussed in \citet{NG9WN} are the jitter and scintillation-noise terms; the latter is separate from the S/N change due to scintillation. Both cause stochastic deviations to the observed pulse shapes such that the matched-filtering assumption of template fitting no longer applies. Jitter, due to single-pulse stochasticity, becomes a significant noise contribution for well-timed, high-S/N MSPs \citep{NG12WN}. \citet{sod+14} measured the timing uncertainty due to jitter for an hour-long observation to be $\sigma_{\rm J} = 48.0 \pm 0.6$~ns, significantly less than the template-fitting errors shown in \citet{IAR}. In addition, since the rms jitter scales inversely with the square root of the number of pulses, this contribution to the TOA uncertainty for a 3 hr 40 min observation will be even smaller in IAR data, and therefore can be ignored. 

Scintillation noise is caused by the finite-scintle effect and results in pulse-shape stochasticity due to an imperfectly known {\it pulse broadening function} \citep{cwd+90,cs2010,Coles+2010}. Its maximum value is approximately the scattering timescale $\taud$, which is inversely proportional to $\Dnud$ by \citep{cr1998, cs2010}
\be 
\taud = \frac{C_1}{2\pi\Dnud},
\ee
where $C_1 \approx 1$ is a coefficient that depends on the geometry of the intervening medium. Since scattering tends to be more significant at higher DMs \citep{Bhat+2004}, we already expect $\taud$ to be small for \PSR. For the minimum value of $\Dnud$ above, we find that maximum value of $\taud \lesssim 1$~ns, and therefore scintillation noise can also be ignored for this pulsar.

\subsubsection{Dispersion Measure Estimation}

Estimation of DM and the subsequent correction of dispersion is critical for proper precision timing. Since the scattering timescale is small, the dominant component to DM estimation errors is that due to the white noise described above \citep{optimalfreq}. For measurements taken at two frequencies $\nu_1$ and $\nu_2$, with the frequency ratio defined as $r \equiv \nu_2 / \nu_1 \ge 1$, the timing uncertainty due to DM estimation is \citep[see Appendix A of][]{DMnu_response}
\be 
\sigma_{\DMhat} = \frac{\sqrt{\sigma_{\nu_1}^2 + r^4 \sigma_{\nu_2}^2}}{r^2 - 1},
\label{eq:DMerr}
\ee
where the $\sigma_\nu$ values are the timing uncertainties at each frequency. Both \citet{cs2010} and \citet{optimalfreq} describe the matrix formalism for calculating this uncertainty when multi-channel measurements are available. Estimating the DM with measurements taken at multiple frequencies results in reduced uncertainties even if the covered range $r$ is the same.

If DM variations are not accounted for in a timing model, as in \citet{IAR}, then additional uncertainties arise. Many types of effects give rise to both stochastic and systematic variations in DM \citep{DMt}. For example, the Earth-pulsar line of sight passing close to the Sun will cause an increase in the DM from the increased electron density of the solar wind \citep{You+07b,NG11SW,Tiburzi+2019}; while the motion through this electron density profile gives rise to systematic variations, any fluctuations in the solar wind will cause variations that are random in nature. The turbulent interstellar medium will also cause rise to DM fluctuations that are random though correlated over time \citep{fc90,pw91}.

The turbulent medium is typically parameterized by a power-law electron density wavenumber spectrum over many orders of magnitude \citep{Armstrong+1995}. An equivalent formulation that one can derive from this spectrum is the DM structure function \citep[e.g.,][]{DMt}, given by 
\be 
D_{\DM}(\tau) \equiv \langle \left[\DM(t+\tau)-\DM(t)\right]^2\rangle,
\ee
where $\tau$ is the time lag and the brackets denote the ensemble average. The structure function of the timing perturbations $\delta t$ is related to the DM structure function by 
\be 
D_{\dt}(\tau) = K^2 \nu^{-4} D_\DM(\tau),
\ee
where $K \approx 4.149 \times 10^3~\mu$s~GHz$^2$~pc$^{-1}$~cm$^3$ is the dispersion constant in observationally convenient units, i.e., for a DM of 1~pc~cm$^{-3}$ and radio emission at 1~GHz, we expect 4149 ~$\mu$s of delay. This is a significantly larger delay than the achievable timing precision since DM {\it estimates} are known much more precisely and therefore the delay can be corrected for to high accuracy.

We can directly relate the rms timing fluctuations for a given time lag to the structure function by
\be
\sigma_\dt(\tau) = \left[\frac{1}{2} D_\dt(\tau) \right]^{1/2}.
\label{eq:sigma_tau}
\ee
For a Kolmogorov turbulent spectrum, we can write the structure function in the power-law form \citep{DMt}
\be 
D_\dt(\tau) = \frac{1}{(2 \pi \nu)^2} \left(\frac{\tau}{\Dtd}\right)^{5/3} = 0.0253~\mathrm{ns}^2~\nu_{\rm GHz}^{-2}~\left(\frac{\tau}{\Dtd}\right)^{5/3}.
\ee
For \PSR, we will first consider the middle value of $\Dtd$ provided by \citet{Keith+2013}, with $\Dtd = 2290$~s.
On the timescale of the $\tau \approx$100 days as shown in \citet{IAR}, the rms timing perturbations for unaccounted for DM variations would be approximately 110~ns at 1.4~GHz and so should be a negligible part of their error budget shown. Nonetheless, if consider this error for longer timespans, as for 1000 and 10000 days, the rms perturbations are 740 ns and 5.0~$\mu$s, respectively\footnote{\added{Again} note that \replaced{this is not the total rms variations here, only those on the timescale of $\tau$ but find the estimate instructive nonetheless.}{Eq.~\ref{eq:sigma_tau} describes the rms fluctuations over a timescale $\tau$ rather than the full rms, which} can be obtained by relating the structure function to the power spectrum of the timing perturbations \citep{DMt} and integrating to find the total variance.}. \added{Note that the predicted structure function extrapolated from $\Dtd$ as given above, using the value of $\Dtd$ estimated from scintillation measurements, is actually lower than what is empirically estimated from the DM timeseries by \citet{Keith+2013} by a factor of 5. This could potentially arise from additional structures in the interstellar medium unrelated to general turbulence \citep{DMt}.}

\subsubsection{Additional Sources of Uncertainty}

\citet{IAR} discuss the RFI environment around the Observatory, including the amplitude and approximate impact on the pulsar observations. While the effect appears to be minimal, RFI is a growing concern at all observatories worldwide. IAR's proximity to the urban center of Buenos Aires means that active real-time mitigation will need to be implemented in the future to reduce any growing impact on the error budget. Also importantly for the IAR Observatory, polarization miscalibration can be a significant contribution to the timing uncertainty, including when both polarizations are measured. \citet{IAR} discuss the development of current and future polarization calibration routines as a goal for the IAR.

In addition to the TOA uncertainty estimates, \PSR is known to show red noise in its timing residuals \citep[e.g.,][]{Kerr+2020}, consistent with rotational spin noise \citep{sc2010,NG9EN}. While red noise will impact the sensitivity towards CWs, the use of matched filtering in detection analyses coupled with the requirement to observe a correlated signal in multiple pulsars means that individual red noise is of less critical significance in CW analyses versus stochastic-background analyses. \added{In other words, while red noise can mimic the low frequency time-correlated signal of the stochastic-background, detectable single sources are necessarily expected to be resolvable from within red noise. In particular at frequencies larger than $\sim$10 nHz steep red noise does not effect CW sensitivity. This was explored in \citet{Xin+2020} where the addition of red noise with a steep spectral index, there in the form of a gravitational wave background, had little effect at frequencies $>$10 nHz. Strong red noise can obviously effect sensitivity, but this is highly dependent on the spectral index. Lastly, at least some of \PSR's red noise appears to be due to time-correlated noise in various receiver-backend combinations at Parkes \citep{Lentati+2016, Goncharov+2021}. Observations from various telescopes allow for the mitigation of these noise sources and more accurate characterization of the intrinsic red noise.}

\subsection{Pulsar GW Sensitivities}

GW detection relies on measuring the signature of TOA perturbations amongst pulsars. In the case of a stochastic background, for instance from the unresolvable sum of signals from SMBHBs, one relies on cross-correlations between these perturbations. For individual SMBHBs, and other single-source signals, one uses a deterministic model. In both cases the sky location and sensitivity of individual pulsars affect the ability of the full PTA to detect gravitational waves. We use the framework discussed in \citet{Hazboun+2019} and \citet{optimalobs} to estimate the sensitivity to single-sources, which we describe here briefly. 

While the ability to claim a detection of GWs from a single SMBHB requires more than one pulsar, the sensitivity to a single source depends on the signal strength and noise in each individual pulsar. These signals are then added up across the network to build a robust detection. There a number of statistics defined in the literature to search for single-source GWs, \citep{Babak+2012,Ellis+2012a,Ellis+2012b,Taylor+2016}. Here we focus on assessing and optimizing the sensitivity of PTAs \citep{optimalobs,Hazboun+2019} given different pulsars and changes in the observing strategies. The framework of \citet[]{Hazboun+2019} uses a match-filter statistic tailored to studying sky-dependent sensitivity. The \snr,  $\rho(\hat k)$, is dependent on the response of an individual pulsar's sky position and noise characteristics but has been averaged over inclination angle and GW polarization
\be
\langle\rho^2(\hat k)\rangle_{\rm inc} = 2 T_{\rm obs}\int_{f_{\rm L}}^{f_{\rm H}} df\>
\sum_i \frac{4}{5}\frac{T_i}{T_{\rm obs}}\frac{S_h(f)}{S_i(f,\hat k)}.\,
\label{e:rho2(k)}
\ee
Here $S_h(f)$ is the strain power spectral density (PSD) of a monochromatic SMBHB signal, $T_i/T_{\rm obs}$ is the fraction of total observation time for the PTA covered by a particular pulsar labeled by $i$, and ${S_i(f,\hat k)}$ is the strain-noise PSD for a particular pulsar. The dependence on sky location, $\hat k$, comes from the quadrupolar respone function of pulsars to GWs. This \snr\ shows that the signals from individual pulsars add independently, but this really only tells part of the story. While the quantity $\rho(\hat k)$ captures the sensitivity of a PTA to individual sources one would need a significant signal in a few pulsars in order to claim a detection.

The PSD ${S_i(f,\hat k)}$ is related to the usual noise spectra, written in \citet{optimalobs} as the sum of white noise, red noise, and the stochastic GW background,
\be 
P_i(f) = P_{\W,i}(f) + P_{\R,i}(f) + P_{\SB}(f),
\label{eq:noise_spectra}
\ee
where the stochastic background term takes a power law form of $P_\SB (f) \propto f^{-\beta}$ with $\beta = 13/3$ for an ensemble of SMBHBs \citep{Jenet+2006}. The PSD ${S_i(f,\hat k)}$ also takes into account the timing model fit of the pulsar, through the inverse noise-weighted transmission function and the response function, which maps the strain from the GWs to the induced residuals in the pulsar's TOAs. See \citet[]{Hazboun+2019} for more details. For uniform white noise given by $\sigma_{\W,i}$ between observations with cadence $c_i$, the white-noise term is simply $P_{\W,i}(f) = 2 \sigma_{\W,i}^2/c_i$. This form demonstrates the strength of the IAR Observatory in observing \PSR: the cadence can be a factor of $\sim$30 larger than for other telescopes, thus significantly reducing the white noise term.

The strain power spectrum for a single CW source with GW frequency $f_0$ and amplitude $h_0$ is \citep{tr2013,optimalobs,Hazboun+2019}
\be 
S_h(f) \equiv \frac{1}{2}\,{h_0^2}\left[\delta_T(f-f_0) + \delta_T(f+f_0)\right]\,.
\label{e:Sh(f)_CW}
\ee
where $\delta_T(f)$ is the finite-time approximating function of a Dirac delta function,
\be 
\delta_T(f) \equiv \frac{\sin(\pi f T)}{\pi f}, \quad \lim_{T \to \infty} \delta_T(f) = \delta(f).
\ee
That is, given the simple CW approximation often used in PTA analyses of an unchanging orbital frequency, over an infinite time, the power spectrum would be a delta function at $f_0$. While more realistic signals, such as those that include a phase shift from the delayed pulsar term and frequency evolution, are usually used in real PTA analyses \citep[e.g.,][]{NG11CW}, the sensitivity is accurately estimated with a circular model. Additionally, it should be noted that this framework includes the signal from the time-delayed pulsar term. An analysis using the pulsar term requires very accurate pulsar distances in order to gain an appreciable signal; perfect knowledge of the distance boosting the squared \snr\ by a factor of $2$. While standard analyses \citep[]{NG11CW} use the pulsar term, it is unclear how much is gained by its inclusion. This has no effect on the ratios of detection thresholds examined here since the factor (of $\sqrt{2}$) for thresholds would cancel.

\section{Current IAR Sensitivity of \PSR}
\label{sec:current_obs}

Here we describe the current observational status of \PSR by the IAR. Observational and derived parameters for \PSR are provided in Table~\ref{table:J0437}. \citet{IAR} demonstrated 88 high-precision timing measurements from A1 and 106 from A2 in its preliminary timing campaign in 2019.

We will assume a standardized phase bin resolution of $\Nphi = 512$. This choice implies that comparisons of pulse S/N will vary between observing setups (NANOGrav uses a uniform 2048 but other groups use variable values) though the TOA uncertainty will still be constant \citep{NG9WN}. A uniform definition of pulse S/N, as in Eq.~\ref{eq:TFerror}, will however allow us to project the sensitivity of the IAR observations more easily and therefore it is still useful to convert between the two quantities $\sigma_{\rm S/N}$ and $S$.

Using the smoothed S/N-weighted average 20-cm template from the Parkes Pulsar Timing Array \citep[PPTA; ][]{Kerr+2020}, we calculated the effective width $\Weff$ of \PSR to be 667.6~$\mu$s (see Table~\ref{table:J0437}). From the residuals in \citet{IAR}, we calculated the median template-fitting uncertainties to be 200 and 325 ns for A1 and A2, respectively. With $\Nphi = 512$, the equivalent median S/N for A1 is 148.8, while for A2 is 90.6. \deleted{While different astrophysical and terrestrial effects will change the observed S/N of the pulses from epoch to epoch,} \added{These values are consistent to those derived in \citet{IAR2}, who analyse an extended IAR data set. The differences in the S/N between both telescopes is not fully explained by the gain differences from dish materials nor differences in observation lengths. Insertion losses are slightly higher for A2, which may additionally account for some of the S/N difference. } We use \replaced{these parameters}{the parameter values given above} to form the basis of our extrapolation to future observing configurations in the next section, \added{noting that they may be conservative estimates such that the performance of A2 on the frontend may improve in the future more than our naive scalings would suggest.}

\section{Extrapolation of IAR Observations of \PSR to Future Systems}
\label{sec:future_obs}

In this section, we provide estimates of the IAR's sensitivity with changes in the observing configuration, namely the use of different receiver ranges for the two antennas. We also compare with sensitivity estimates of other facilities observing \PSR and show that the IAR's substantial cadence, i.e., its time on the source, will provide it with a \replaced{unique}{special} capability in observing this pulsar in the future.

\subsection{Upgraded Observing Configurations}
\label{sec:upgraded}

To estimate the TOA uncertainties for each system, we combined the various noise components in the following manner. We \replaced{took}{scaled} the A1 and A2 median S/Ns as estimated in \S\ref{sec:current_obs} and scale by the ratio of the flux densities from 1400 MHz to the new center frequency, the ratio of the square root of the bandwidths and number of polarization channels, and inversely by the ratio of the receiver temperatures. With the new S/N, we calculated a new template-fitting error for the receiver. With the two center frequencies, we were able to calculate the TOA uncertainty due to DM estimation (Eq.~\ref{eq:DMerr}); we assumed that in all future timing analyses, DM will be estimated on a time-varying basis such that the rms timing fluctuations are negligible (e.g., from Eq.~\ref{eq:sigma_tau}), otherwise the unmodeled DM variations will make any high-precision timing experiment impossible. Lastly, we added in 100 ns of uncertainty due to polarization miscalibration, comparable to the value calculated by \citet{vS2006}. 

Even though we approximated the observations across each band as single measurements, ignoring the varying spectral index with the band, changes in pulse width, etc., for simplicity we calculated the TOA uncertainties as if there are two ``spot'' measurements at each center frequency. While this is a simplification, it provided us with a {\it conservative} estimate since a full least-squares fit formalism will lead to reduced TOA uncertainties.

In summary, our TOA uncertainties consisted of template-fitting errors, DM estimation errors, and polarization miscalibration errors, added together in quadrature.

We considered several different configurations: both the current and future setups as described in \citet{IAR}, and setups where additional receivers are built. We restricted ourselves to discussing the configurations broadly, assuming that the receiver frontend is matched with a backend that will be able to adequately sample and coherently dedisperse the data. \cite{IAR} describe the future planned upgrades to the antennas, which include receivers reaching system temperatures $T_{\rm sys} < 50$~K, a 500~MHz bandwidth in a range between 1 and 2 GHz, and a new FPGA-based CASPER board \added{\citep{Hickish+2016}} backend capable of processing the increased bandwidth. For other configurations we list, we assumed similar matching as we primarily wanted to focus on estimating the TOA uncertainties given the assumed development of a specific system. We estimated these uncertainties under two scenarios with the current equipment, three in which one or more of these higher-bandwidth systems is deployed, and one optimistic scenario in which ultrawideband receivers are deployed on both antennas. The full list of assumptions is provided in Table~\ref{table:configs} along with the TOA uncertainties we estimate. The gains of the antennas in our analysis did not vary though some dish or efficiency improvements may be possible in the future, \added{ for example, resurfacing the dishes in the future may improve the aperture efficiencies slightly as compared with their current materials \citep{Testori+2001,IAR}}.

We describe the rationales for considering each configuration below.

\begin{enumerate}[label=C\arabic*:, topsep=-11pt, itemsep=0pt]
    \item Current. This configuration is as described in \citet{IAR}\added{, see their Table 1}.
    \item Optimized. Since the center frequencies are tunable, we allowed for the maximum separation in frequencies to minimize TOA uncertainties from DM misestimation. We also halved A1's bandwidth coverage in favor of dual-polarization measurements  \added{(see their Section 2.1 regarding these two modes for A1)}, a requirement for high-precision timing.
    \item Wideband. With a modest upgrade of two receivers and backends covering 500~MHz each as discussed in \citet{IAR}, the two antennas can cover the 1-2 GHz range. The target system temperature is $T_{\rm sys} < 50$~K. While some parts of the band will be lost due to RFI, no significant segments of the band are currently lost across this frequency range \citep{IAR}.  
    \item Low Frequency. Instead of two receivers covering the full L-band range, we instead selected one 500~MHz receiver to cover the top end of L band (1.5-2.0 GHz) for A1 and then for A2 considered a receiver from 400-450 MHz. This lower frequency range is used by NANOGrav at Arecibo for some pulsars. Its primary allocation in Argentina\footnote{Ente Nacional de Comunicaciones (ENACOM) follows ITU-R regulations, as described in \citet{ENACOM}.} is for maritime communication and radionavigation  (see \citealt{ITU} for Region 2) which helps to limit fixed sources of RFI. In addition, scattering will minimally affect \PSR given its low DM, and so the increased flux density (see Table~\ref{table:J0437}) can lead to high-S/N TOAs, also providing a significant frequency difference to estimate DM. For the L-band receiver, we used the higher end of the possible bandwidth range to minimize TOA uncertainties due to DM misestimation.
    \item High Frequency. As a parallel to C4, we considered instead using a second receiver at higher frequencies, in the 2.5 GHz (S band) range which is also used by NANOGrav at Arecibo for some pulsars. This region of the frequency spectrum often contains significant RFI due to overlap with wireless communications. Nonetheless, we considered the potential for such a system. For the L-band receiver, we used the lower end of the possible bandwidth range to minimize TOA uncertainties due to DM misestimation.
    \item Dual Ultrawideband. As an optimistic setup, we considered the receiver systems as described in \citet{DSA2000} for the DSA-2000. This observatory will employ low-cost receiver systems from 0.4-2.0~GHz for each of its 2000 antennas. While their target system temperature is 25 K, we kept 50 K for use in this considered configuration. Since the receiver setup is identical for both antennas, for simplicity we assumed two bands centered at 700 and 1500 MHz with 600 and 1000 MHz of bandwidth, respectively, to cover the 400-2000 MHz range, and included a factor of $\sqrt{2}$ in our calculation.
\end{enumerate}~

As discussed in \S\ref{sec:shorttimenoise}, one astrophysical cause of the S/N changing is due to diffractive scintillation. In considering wider bandwidths, the number of scintles will grow, thus causing the pulse S/N to tend more towards a mean value rather than cover an exponentially-distributed range. In practice, this means that for low-bandwidth setups as in the current configurations, there are epochs in which the template-fitting error in Eq.~\ref{eq:TFerror} breaks down \citep{NG9} and TOAs cannot be reliably estimated, resulting in a loss of usable data. In addition to the mean S/N $S_0$ increasing due to wider bandwidths, the change in the diffractive scintillation PDF means that the TOAs will tend towards higher mean/median values, without a loss of data. Thus, we considered the extrapolation from the median S/N in \S\ref{sec:current_obs} to be more robust.

For our analysis, we assumed that configurations C3 through C6 are operational by 2023, a rapid timeline, and assumed that the observing conditions are static into the future. As with all facilities, future upgrades can yield additional sensitivity. In addition, growing RFI will impact timing sensitivity as well. 
\citet{IAR} show that $\sim 10$\% of observing times \added{are significantly affected due to RFI}, with significantly less time at night. RFI affects both antennas differently given their differing proximities to the local IAR offices. The RFI tends to be narrowband and thus can be excised while retaining the majority of the band. In our work, we ignored the role of RFI, noting that it will play an important but small contribution in the assumptions we are making regardless.

\begin{deluxetable*}{l||cc|cc||cc|cc|cc||cc}
\tablecolumns{13}
\tablecaption{Observing Configurations for the IAR Antennas}
\tablehead{
\colhead{Parameter} & \multicolumn{2}{c}{C1: Current} & \multicolumn{2}{c}{C2: Optimized} & \multicolumn{2}{c}{C3: Wideband} & \multicolumn{2}{c}{C4: Low Freq} & \multicolumn{2}{c}{C5: High Freq} & \multicolumn{2}{c}{C6: Dual UWB}\\
& \colhead{A1} & \colhead{A2} & \colhead{A1} & \colhead{A2} & \colhead{A1} & \colhead{A2} & \colhead{A1} & \colhead{A2} & \colhead{A1} & \colhead{A2} & \colhead{A1} & \colhead{A2}
}
\startdata
Circular Polarizations ($N_{\rm pol}$) & 1 & 2 & 2 & 2 & 2 & 2 & 2 & 2 & 2 & 2 & 2 & 2\\
Center Frequency ($\nu_0$, MHz) & 1400 & 1400 & 1128 & 1572 & 1750 & 1250 & 1750 & 425 & 2500 & 1250 & 1200 & 1200\\
Bandwidth ($B$, MHz) & 112 & 56 & 56 & 56 & 500 & 500 & 500 & 50 & 500 & 500 & 1600 & 1600\\
Receiver Temperature ($T_{\rm rcvr}$, K) & 100 & 110 & 100 & 110 & 50 & 50 & 50 & 50 & 50 & 50 & 50 & 50\\
Gain ($G$, K/Jy) & 0.084 & 0.077 & 0.084 & 0.077 & 0.084 & 0.077 & 0.084 & 0.077 & 0.084 & 0.077 & 0.084 & 0.077 \\
\hline 
\parbox{3cm}{\vspace{2pt}\PSR TOA \newline uncertainty ($\sigma_{\rm eff}$, $\mu$s)\vspace{2pt}} & \multicolumn{2}{c|}{2.95} & \multicolumn{2}{c||}{2.31} & \multicolumn{2}{c|}{0.46} & \multicolumn{2}{c|}{0.34} & \multicolumn{2}{c||}{0.36} & \multicolumn{2}{c}{0.19}\\
\enddata
\tablecomments{Current (C1) configuration values are taken from \citet{IAR}\added{, see their Table 1}. We have defined the system gain in terms of K/Jy, the ``forward'' gain, with values of $1/G$ instead given in \citet{IAR} and Table~\ref{table:J0437}. The maximum tracking time ($T_{\rm obs}$) for both antennas is 3 hr 40 min\added{, and we assume observations are performed at a daily cadence}.}
\label{table:configs}
\end{deluxetable*}

\subsection{Comparison with Other Observatories}

\begin{figure}[t!]
\hspace{-10pt}
\includegraphics[width=0.52\textwidth]{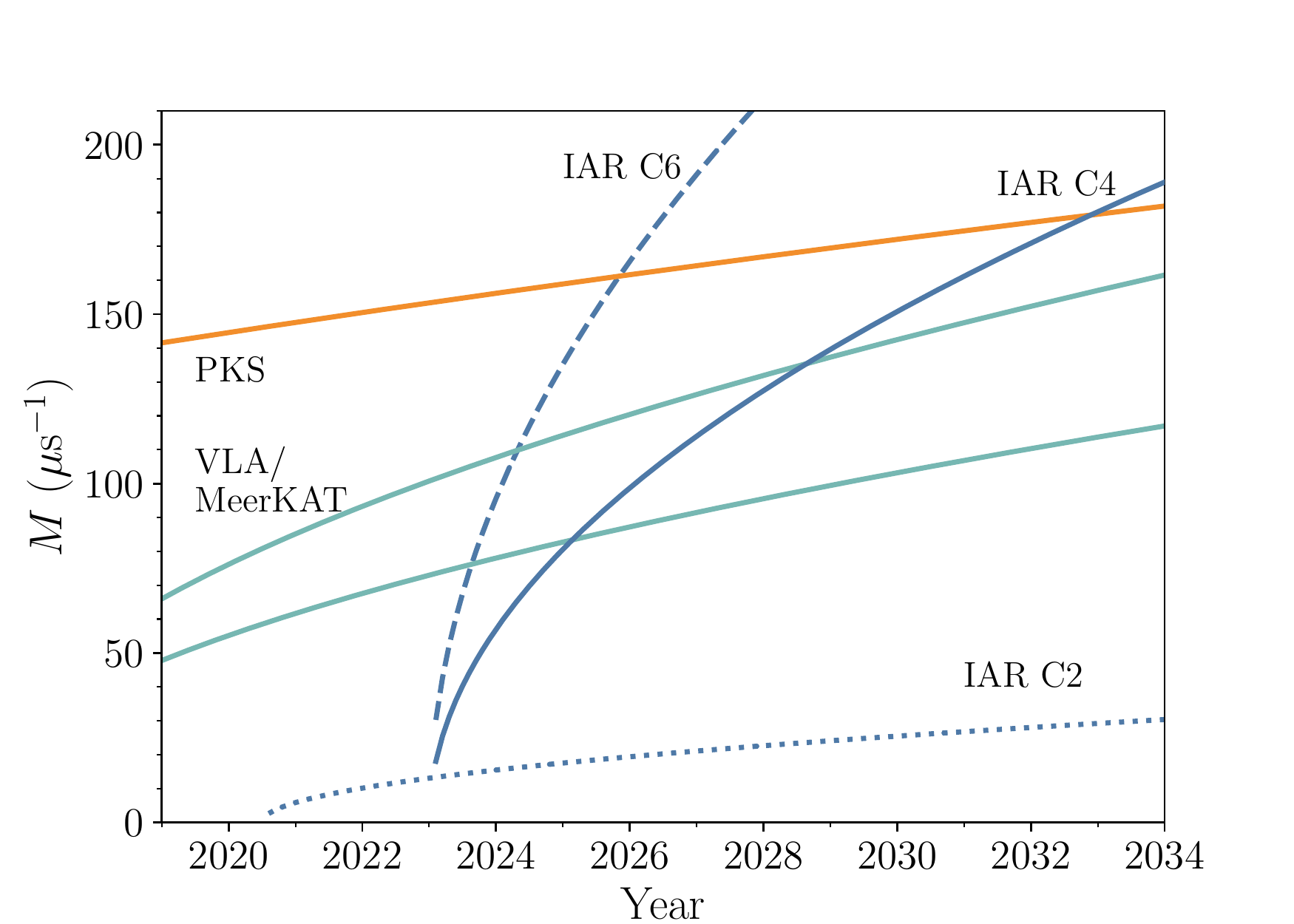}
\caption{Plots of the timing precision metric $M$ (see Eq.~\ref{eq:M}) versus time. Increased $M$ implies improved sensitivity. The blue lines denote the various IAR configurations, where C4 and C6 start in 2023. The metric for Parkes is shown in orange and the two teal curves show the metric for the VLA\added{/MeerKAT} assuming two different estimates for polarization noise.}
\label{fig:config_comp}
\end{figure}

Using the single-pulsar-dependent terms in Eq.~\ref{e:rho2(k)} and the instrumental/telescope-dependent components of the TOA uncertainty, defined below, we constructed a metric that describes the overall timing precision of a pulsar in an array (and dropping the subscript $i$):
\be 
M \equiv \left(\frac{Tc}{\sigma_{\rm eff}^2}\right)^{1/2} = 34.6~\mu\mathrm{s}^{-1}~\left(\frac{T}{10~\mathrm{yr}}\right)^{1/2}\left(\frac{c}{12~\mathrm{yr}^{-1}}\right)^{1/2}\left(\frac{\sigma_{\rm eff}}{0.1~\mu s}\right)^{-1}.
\label{eq:M}
\ee
When $T$ is the total observing baseline in years, $c$ is the observing cadence in years$^{-1}$, and $\sigma_{\rm eff}$ is the \added{{\it effective}} telescope-dependent \deleted{{\it stochastic}} TOA uncertainty in $\mu$s, then $M_i$ has units of $\mu$s$^{-1}$. We define $\sigma_{\rm eff}$ here as the combination (quadrature sum) of the short-timescale white noise, the DM estimation uncertainty due to the white noise (Eq.~\ref{eq:DMerr}), and polarization calibration uncertainty (taken to be 100~ns, see \S\ref{sec:upgraded}. The short-timescale white noise includes template-fitting, jitter, and scintillation-noise uncertainties. As discussed in \S\ref{sec:sensitivity}, jitter becomes negligible for the duration of observations conducted by the IAR but not necessarily for other observatories. Scintillation noise is negligible given the small scattering timescale.

The quantity $M$ acts like a signal-to-noise ratio, where larger $M$ implies a higher sensitivity, though we note that it is not a direct relationship. \deleted{As the number of observations $Tc$ increases, then $M$ increases as the square root of that number, and as $\sigma_{\rm eff}$ decreases, then $M$ increases linearly just as many signal-to-noise ratios behave.} Alternatively, this quantity can be viewed as proportional to the (square) inverse of the error on the mean of the residuals.

Figure~\ref{fig:config_comp} shows curves of $M$ as a function of time for a number of different telescopes and configurations. The three blue curves denote three different IAR configurations as listed in Table~\ref{table:configs}, where for C2 we take the start time as mid-2020 and for all other upgraded configurations we assume an aggressive program start time of January 2023. Delays will shift the curves to the right. Nonetheless, it is quite clear how drastically instrumentation improvements can improve $M$.

Comparisons to other observatories are described in the following subsections.

\subsubsection{The Parkes Telescope}

The PPTA has observed \PSR since 1996 \citep{Hobbs2013}, with a typical cadence of once every two weeks in three bands \citep{Manchester+2013,Reardon+2016,Kerr+2020} for approximately one hour in each of the 20 (1400 MHz) and 10/50 cm (3100/730 MHz) observations \citep{Hobbs2013}. The backend systems have improved over time, resulting in many different noise contributions to consider. We estimated the equivalent $\sigma_{\rm eff}$ as follows. Using the PPTA Second Data Release\footnote{Files are available at \url{https://doi.org/10.25919/5db90a8bdeb59}.} \citep{Kerr+2020}, we used the empirically-derived noise parameters as an estimate for the telescope-dependent \deleted{stochastic terms} \added{$\sigma_{\rm eff}$} described previously.

First, we took parameters that describe modifications to the template-fitting uncertainties, namely a noise added in quadrature (EQUAD) with the resultant quantity multiplied by a scaling factor (EFAC), to provide us with an estimate of the white noise for each TOA \citep{Reardon+2016}. Since the white noise parameters in \citet{Kerr+2020} were derived from sub-banded TOAs, i.e., those taken from small frequency channels over a wider bandwidth, we then computed the epoch averaged error as in \citet{optimalfreq} (and also see references therein) to provide us with the amount of white noise per epoch per band. This amounts to computing the square root of $(\mathbf{U}^T \mathbf{C}^{-1} \mathbf{U})^{-1}$, where $\mathbf{U}$ is a column matrix of ones and $\mathbf{C}$ is the covariance matrix, which in this case is a diagonal matrix containing the squared of the modified TOA uncertainties.

These epoch-averaged TOA uncertainties describe the measured white noise in the \citet{Kerr+2020} data set. We also included the measured ``band noise'' terms, which describe additional noise as a function of specific frequency ranges. Such band noise can describe either additional chromatic propagation effects or terrestrial effects such as from RFI or polarization miscalibration \citep{Lentati+2016}. Since in \S\ref{sec:sensitivity} we argued that \PSR\ likely has low levels of noise due to unmodeled propagation effects such as scattering, then we took the band noise to be due to terrestrial effects and as such consider it an important contribution to $\sigma_{\rm eff}$. \added{To add a band noise component to the TOA uncertainties, we determined the variance numerically by integrating the power-law spectrum form described by the amplitude and index parameters (values set by the parameter TNBandNoise in the data files). These variances were added to the white noise variance per epoch per band computed above to appropriately adjust the uncertainties.} Finally, to use a single number to project $M$ to, we computed the average ``weight'' of each uncertainty ($w_k = 1/\sigma_k^2$ for the $k$-th TOA), computed the mean weight, and then calculated the square root of reciprocal of this mean weight \added{- our end result was $\sigma_{\rm eff} = 166$~ns}. \added{In our sensitivity extrapolation calculation, we did not account for any remaining variance in the timing residuals, e.g., due to DM estimation or improper polarization calibration, which we assumed in our analysis was accounted for elsewhere in the original noise modeling of the data set.}

We note that an ultrawideband receiver system has been deployed at Parkes \citep{Hobbs+2020}, which may result in reduced uncertainties in the future if RFI does not become sufficiently problematic. If so, this will improve future estimates of $M$ for Parkes as compared with what is shown in Figure~\ref{fig:config_comp}. However, the majority of the amplitude of $M$ comes from the large observing baseline $T$ more so than the TOA uncertainty, so it is unclear by how much $M$ will be affected.

\subsubsection{The Very Large Array}

NANOGrav has also begun\footnote{Proposal IDs VLA/16B-240,  VLA/18A-210, VLA/19A-356} a program to observe \PSR with the VLA monthly between 1 and 4 GHz, with 30 total minutes devoted to L band (1-2 GHz) and 30 minutes devoted to S band (2-4 GHz). While the full hour includes overhead time, for the purposes of our calculations, we will assume each observation is 30 minutes long.

Without empirically-derived white-noise parameters as in the PPTA data release, while we could work out a full calculation of the white noise plus DM uncertainties, we will instead provide a heuristic argument to demonstrate that the observational setup with the VLA will result in a lower $M$. Let us assume that jitter is the only white-noise contribution to the TOA uncertainty. Using the values in \citet{sod+14} (see also Table~\ref{table:J0437}), we have that the rms jitter for a 1 hour observation is 48~ns at L band and 41 ns at S band. As jitter scales as the square root of the number of pulses, the 30-minute rms jitter will be 68 and 58~ns for the two bands, respectively. For DM estimation errors, we will assume for this argument that the two spot frequencies are 1 and 4 GHz rather than center frequencies of 1.5 and 3 GHz, which will underestimate the uncertainty but will not matter for a comparison. Using our assumed numbers, calculating $\sigma_{\DMhat}$ using Eq.~\ref{eq:DMerr}, and then finding the total infinite-frequency TOA uncertainty, we have 76~ns per observation as a {\it minimum} bound. With observations starting in 2016, and assuming 12 observations per year continued until 2033, we have $M$ between 123 and $170~\mu\mathrm{s}^{-1}$, assuming 50 and 100 ns of polarization calibration uncertainty, respectively.

\subsubsection{MeerKAT}

MeerKAT has begun pulsar-timing observations with the MeerTime project \citep{TPA,MeerTime2020}, with one goal of the project to observe MSPs. With the expected next generation of receivers extending the observatory's bandwidth range, in a subarray mode, MeerKAT will be able to observe from 0.9-3.5 GHz \citep{MeerTime2018,MeerTime2020}. If again we assume that the pulsar is jitter-dominated as with the VLA, then we expect comparable numbers as calculated above.

\subsubsection{Comparing $M$ for Different Observatories}

Figure~\ref{fig:config_comp} shows $M$ for the VLA (again, comparable with MeerKAT), Parkes (PKS), and three configurations for the IAR: C2, C4, and C6. Given the assumptions presented above, we see that an aggressive timeline for improved instrumentation will drastically improve IAR's sensitivity to \PSR\ in comparison to other facilities. Delays in this timeline will shift the curve to the right but the conclusion remains the same: a modestly upgraded IAR facility (C4) will drastically aid in single-source GW science right in the era of the first CW sources, as well as the first observations by LISA. Using state-of-the-art yet low-cost instrumentation \citep[e.g., C6 using similar instrumentation as described in][]{DSA2000} will provide the IAR with unparalleled sensitivity to \PSR and thus a unique contributor to GW science with PTAs by the middle-to-end of the current decade. We note that other observatories like Parkes \replaced{will still have an advantage over}{are complementary for} long-timescale ($\sim$decade) GW periods since its observations of \PSR\ cover a sufficient number of cycles whereas observations by the IAR alone will not.

\section{GW Projections for \PSR}

\label{sec:projections}
In order to demonstrate the usefulness of \PSR as an addition to the NANOGrav observing campaign, we projected the status of the PTA described in \citet{NG11} into the next decade. Here we are only interested in how \PSR contributes to the sensitivity for deterministic signals so we simply extended the baselines of the data set most recently used in these GW analyses, using the noise characteristics and sky locations of the current set of pulsars and those estimated for \PSR at IAR. We only investigated single-source sensitivity because 1) the GWB is forecasted to be detected long before the dates of these projections, and 2) the short baseline of \PSR when it is added to NANOGrav data sets will not be very useful for detection/characterization of the GWB until it has sensitivity at lower frequencies. 

Using the Python package {\tt hasasia} \citep{hasasia} we simulated a PTA with the same characteristics as \citet{NG11} but added 17 years of data to use as a baseline for our comparison, ending in 2033. \added{We used the noise models/parameters laid out in \citet{NG11}, including the measured TOA residual rms values, and where significantly measured, red noise. These models estimate various white noise contributions, mismodeled chromatic effects and intrinsic spin noise.} We also simulated two versions of a PTA with IAR observations of \PSR\ from 2023 onwards, i.e., 10 years of observations. \added{Similarly we used a total residual rms value and red noise parameters, given in \citet{Kerr+2020}. In particular we added red noise with $\log_{10}A_{\rm RN}(f_{\rm yr})=-14.0186 $ and spectral index of $\gamma = 3.17$.} The two versions correspond to using configurations C4 and C6.

\begin{figure}[t!]
\includegraphics[width=0.48\textwidth]{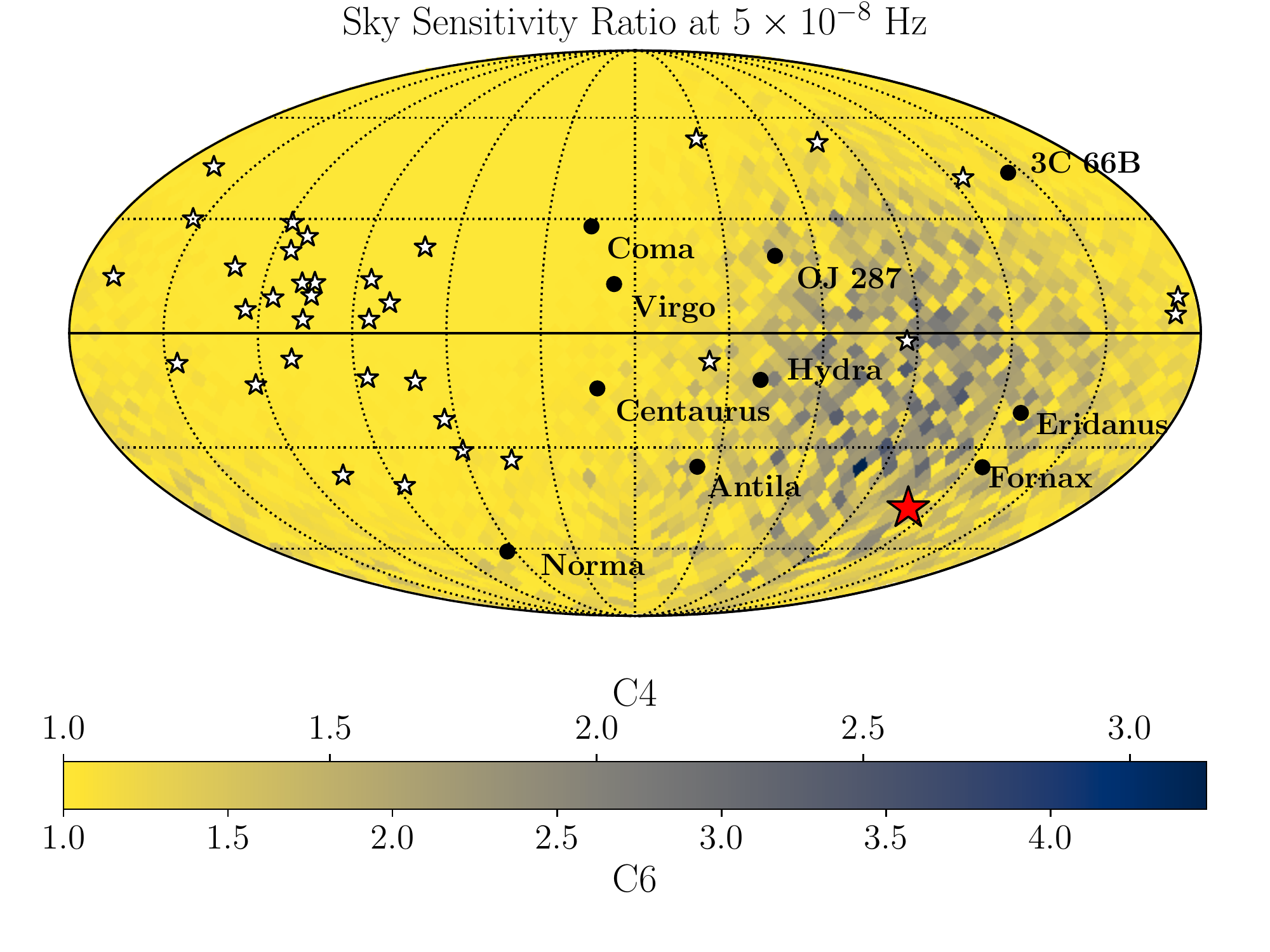}
\caption{Ratio of the GW sensitivity when adding \PSR to the PTA using configurations C4 and C6 versus otherwise. The sensitivity is calculated at \replaced{10$^{-8}$~Hz}{$5 \times 10^{-8}$~Hz}. NANOGrav 11-year data set pulsars are shown in white stars, \PSR in the red star. Possible targets of interest for CW signals are shown in the black dots and labeled. Equatorial coordinates are shown, with right ascension increasing to the left every 2$^{\rm h}$ starting from the right and declination increasing upwards every 30$^\circ$.
The structure of the sky map is the same for C4 and C6 but the corresponding colormap scale is shown at the bottom for each. We see clear significant increases to the sensitivity in NANOGrav's current ``blind spot'' on the sky due to the addition of \PSR.
}
\label{fig:skymap}
\end{figure}

Figure~\ref{fig:skymap} shows the ratio of the sky sensitivity in source strain \citep[see Eq.~80 in ][]{Hazboun+2019} when adding observations of \PSR versus not. We show the positions of the NANOGrav 11-year data set pulsars (white stars), \PSR (red star), and nearby galaxy clusters and individual targets of interest \citep[e.g.,][]{Mingarelli+2017,cz2018,NG11CW,3C66B}. \replaced{We used a fiducial GW frequency of 10$^{-8}$~Hz (periods of $\sim$3~yr), approximately where the current most sensitive CW upper limits are \citep{NG11CW}. The sky maps are identical in structure but the overall sensitivity varies when using C4 versus C6.}{We used a GW frequency of $5\times10^{-8}$~Hz to highlight where the observations of \PSR make a sizable impact on the sensitivity. The sky maps are identical in structure but the overall sensitivity varies when using C4 versus C6. At a frequency of 10$^{-8}$~Hz, approximately where the current most sensitive CW upper limits are \citep{NG11CW}, the increase in source strain is only about $15\%$, while at 10$^{-9}$~Hz any increase in sensitivity is negligible. This dependence on frequency is in part due to the sizable amount of red noise in the TOAs of \PSR.}

We see clear indications that observations with \PSR versus without will increase the sensitivity of the NANOGrav PTA. In the case of C6, by 2033 the array's sensitivity improves by a factor of \replaced{$\approx$2-7}{$\approx$2-4.5} for 20\% of the sky. Even with C4, the improvement ratio is $\approx$1.4-3. Since the observed GW strain is inversely proportional to the distance of a source, a factor of 2 increase in sensitivity leads to the array being able to search out a factor of 2 in distance, indicating that the volume NANOGrav will be able to probe in its current ``blind spot'' will be drastically increased.


\section{Predictions for Observations of Other MSPs}
\label{sec:other_MSPs}

\begin{deluxetable*}{l|cc|ccccc|c}
\tablecolumns{9}
\tablecaption{Predicted Sensitivity of Other Observable MSPs}
\tablehead{
\colhead{PSR} & \colhead{$P$} & \colhead{DM} & \colhead{$W_{50}$} & \colhead{$W_{50}/P$} & \colhead{$W_{\rm eff}$} & \colhead{$\bar{U}_{\rm obs}$} & \colhead{$I_{1400}$} & \colhead{$R$}\\
 & \colhead{(ms)} & \colhead{(pc cm$^{-3}$)} & \colhead{($\mu$s)} & \colhead{(\%)}& \colhead{($\mu$s)} &  & \colhead{(mJy)} & 
\label{table:pulsars}
}
\startdata
J0437$-$4715 & 5.76 & 2.6 & 141.8 & 2.5\% & 667.6 & 0.063 & 158.0 & 1.000\\
J1017$-$7156 & 2.34 & 94.2 & 72.0 & 3.1\% & 292.6 & 0.036 & 0.9 & 0.022\\
J1125$-$6014 & 2.63 & 53.0 & 131.6 & 5.0\% & 257.1 & 0.073 & 0.9 & 0.013\\
J1600$-$3053 & 3.60 & 52.3 & 92.7 & 2.6\% & 469.3 & 0.051 & 2.5 & 0.028\\
J1603$-$7202 & 14.84 & 38.0 & 316.2 & 2.1\% & 1480.7 & 0.060 & 3.5 & 0.010\\
J1643$-$1224 & 4.62 & 62.4 & 318.6 & 6.9\% & 978.7 & 0.098 & 4.7 & 0.013\\
J1730$-$2304 & 8.12 & 9.6 & 573.1 & 7.1\% & 891.5 & 0.109 & 3.8 & 0.010\\
J1744$-$1134 & 4.07 & 3.1 & 137.7 & 3.4\% & 512.2 & 0.037 & 2.5 & 0.035\\
J1909$-$3744 & 2.95 & 10.4 & 43.8 & 1.5\% & 266.2 & 0.017 & 1.7 & 0.100\\
J2241$-$5236 & 2.19 & 11.4 & 64.3 & 2.9\% & 248.3 & 0.033 & 1.8 & 0.058\\
\enddata
\end{deluxetable*}

Our estimates and extrapolations for PSR~J0437$-$4715 can be used to predict the general timing performance of other pulsars observable with the IAR. While PSR~J0437$-$4715 is the brightest known MSP, other factors influence the overall timing precision. Table~\ref{table:pulsars} provides a list of some pulsars observed by the PPTA \citep{Kerr+2020}, which we use as a representative list to quantify potential targets.

For the same observational setup, we expect that the period-averaged signal-to-noise ratio $\bar{S}$ will be proportional to the period-averaged flux density $I_0$ at a fiducial frequency $\nu_0$, using the notation of \citet{optimalfreq}. Wide-bandwidth observations require that the spectral index of the pulsar flux be considered. In general, the minimum TOA uncertainty due to a finite S/N is given by Eq.~\ref{eq:TFerror}, which uses the peak-to-off-pulse S/N, $S$. The conversion between the two is given by $\bar{S} = \bar{U}_{\rm obs}S$, where 
\be 
\bar{U}_{\rm obs} =\frac{1}{\Nphi}\sum_{i=0}^{\Nphi-1} \Uobs(\phi_i)
\ee
is the mean value of the template shape $\Uobs$. A smaller value of $\bar{U}_{\rm obs}$ implies a sharper pulse profile which yields improved TOA uncertainties. Therefore, combining the pulse-shape-dependent factors with the pulsar fluxes, we can relate the ratio of the template-fitting uncertainties between two pulsars A and B as
\be 
R \equiv \frac{\sigma_{\rm S/N, A}}{\sigma_{\rm S/N, B}} = \frac{W_{\rm eff, A}\bar{U}_{\rm obs,A}/I_{\rm 0,A}}{\W_{\rm eff,B}\bar{U}_{\rm obs,B}/I_{\rm 0,B}}.
\label{eq:sigma_ratio}
\ee

With Eq.~\ref{eq:sigma_ratio}, we were able to calculate the template-fitting errors for the pulsars provided in Table~\ref{table:pulsars} compared to PSR~J0437$-$4715. In the table, the larger $R$ is, the better the template-fitting uncertainty will be. This analysis does not take into account other sources of uncertainty in TOA estimation.  For example, sources with higher DM are more likely to be affected by pulse broadening due to interstellar scattering \citep{Bhat+2004}. \added{In addition, Eq.~\ref{eq:sigma_ratio} also does not take into account the declination-dependence of the tracking times of the antennas, which will introduce a $T_{\rm obs}$-dependence affecting the S/N for each pulsar.}

We can more intuitively understand the impact of the pulse shape metrics compared to the flux densities in Eq.~\ref{eq:sigma_ratio} by considering a  Gaussian-shaped pulse. The effective width $\Weff \propto (W_{50}P)^{1/2}$, the geometric mean of the pulse full-width-at-half-maximum $W_{50}$ and the pulse period $P$ (following from Eq.~\ref{eq:Weff}, see also \citealt{NG9WN}). In considering a Gaussian pulse, the mean value of the pulse will be the integral of the pulse shape divided by the pulse period, yielding the approximate relationship $\bar{U}_{\rm obs} \propto W_{50}/P$ for sufficiently narrow pulses. Combining, we have
\be 
\Weff \bar{U}_{\rm obs} \propto \left(W_{50}P\right)^{1/2} \frac{W_{50}}{P} = P \left(\frac{W_{50}}{P}\right)^{3/2}.
\label{eq:proportionalities}
\ee
We write the second equality in terms of the pulse duty cycle, $W_{50}/P$. We see from Eq.~\ref{eq:proportionalities} that decreases in either the pulse period or the duty cycle will lead to a lower   $\sigma_{\rm S/N}$, which is to be expected. These improvements can help compensate for a difference between the flux densities of pulsars. A factor of two decrease in both the pulse period and duty cycle can then offset a factor of five decrease in the flux density of one pulsar compared to another and lead to equivalent $\sigma_{\rm S/N}$. Note that in comparison to \PSR, decreases in $P$ or $W_{50}/P$ by more than a factor of $\sim$2 are generally not observed.

Table~\ref{table:pulsars} gives the various pulse shape parameters for ten pulsars observed by the PPTA in the declination range of IAR and with $R > 0.01$. Pulse shape metrics were measured using the 20-cm Parkes Digital Filterbank System 4 (PDFB4) templates\footnote{Accessible at \url{https://doi.org/10.25919/5db90a8bdeb59}.} in \citet{Kerr+2020}. We used the median flux density at 1400 MHz reported by \citet{Kerr+2020} in our analysis; the scintillation properties of the different pulsars will affect $R$ but we ignored this here and only considered a typical observation. From these measurements, we calculated $R$ according to Eq.~\ref{eq:sigma_ratio}.

Another pulsar with one of the lowest DMs, PSR~J1744$-$1134, has been detected by the IAR in individual observations (L.~Combi, private communication). As with \PSR, we can then consider the impact of daily cadence using the IAR compared with $\sim$monthly observations available at other facilities. Let us naively consider \PSR observed with $\sigma_{\rm eff} \approx \sigma_{\rm S/N} = 200$~ns. While $R = 0.035$ for PSR~J1744$-$1134, leading to many microseconds of uncertainty per epoch, the monthly-averaged uncertainty reduces by a factor of $\sqrt{30}$ to approximately 1~$\mu$s. This argument ignores differences in spectral index\added{, differences in $T_{\rm obs}$ due to the declination,} and other frequency-dependent profile changes but serves to demonstrate the potential of IAR's \deleted{unique} capabilities. Observations of PSR~J1744$-$1134 with daily cadence will be comparably as sensitive as other pulsars with microsecond timing observed with a monthly cadence, making observations of the pulsar of sufficient quality for PTA science. Several other possible targets, such as PSRs~J1909$-$3744, J2241$-$5236, and possibly also J1600$-$3053, may be suitable candidates for daily monitoring.

\section{Concluding Remarks}
\label{sec:conclusion}

The IAR's \replaced{unique}{special} capabilities to observe \PSR will lead to a significant increase in CW sensitivity for PTAs. \added{Several other 30-m-class telescopes exist in the Southern Hemisphere: for example, the Hartebeesthoek Radio Astronomy Observatory \citep[HartRAO;][]{HartRAO} and Mount Pleasant Radio Observatory \citep{MtPleasant} have already been used for long-term pulsar observations, and in Argentina the European Space Agency's Deep Space Antenna 3 tracking station \citep{DSA3} and the joint Argentina-China deep space network antenna CLTC-CONAE-NEUQU\'{E}N \citep{NEUQUEN} will complement the IAR with dedicated time for national radio astronomy projects. In all cases, significant improvements to the instrumentation, e.g., increased bandwidths, would be required for obtaining high-precision TOA estimates of PSR~J0437$-$4715. The IAR is currently specifically targeting high-precision pulsar timing as a goal science driver for its observatory.} However, it is vital that the \added{IAR} observatory receives the necessary upgrades to make this a possibility; without dual-polarization observations at frequencies spaced widely enough to accurately estimate DM, the IAR will not be able to meet the target sensitivities described here. This will require careful polarization calibration and the RFI environment around the observatory must remain in a clean state.

In addition to direct CW sensitivity, high-cadence observations can contribute to other PTA-related science. For example, daily monitoring of DMs and scintillation which will feed back into understanding pulsar timing noise models is a planned goal of the IAR \citep{IAR}. Such observations will complement those of Northern-Hemisphere facilities like CHIME \citep[$\delta >-20^\circ$\added{;}][]{Ng2018} which are beginning to provide unprecedented measurements of the interstellar medium on short timescales \citep{Ng+2020}. For example, given the proper motion of \PSR of 141~mas/yr \citep{Kerr+2020} and distance of 156.8 pc, the pulsar's transverse velocity is 105~km/s. Over 10 years of daily observations, the line of sight will probe length scales in the interstellar medium between 0.06 -- 200~AU. \deleted{Considering the relatively long tracks on \PSR, breaking the observations into individual hour-long measurements would yield a factor of $\sim$2 reduction in the DM precision but the ability to probe scales down to 0.0025 AU, or half of a solar radius.}

International PTA efforts will also benefit from the inclusion of IAR observations. The International Pulsar Timing Array (IPTA) collaboration helps to coordinate these worldwide efforts and while the potential target pulsars listed here would not be new additions to any future combined data set, such a data set will have improved sensitivity to GWs of all kinds with additional high-precision data. \added{While Figure~\ref{fig:config_comp} shows the C4 or C6 configurations matching the sensitivity of Parkes on the timescale of years to a decade, combination of these data as part of the IPTA would yield the optimal results. High-cadence observations obtained by the IAR would complement the longer-term observations by Parkes to probe across GW frequencies. In addition with the added sensitivities of other facilities performing new observations of PSR~\J and the sky coverage of all of the pulsars in the remainder of the combined IPTA data set, we can expect the IAR to be a significant contributor to GW science within even the next several years. In addition,}  uniform observations of several pulsars by the IAR can additionally help constrain systematic variations between different frontend/backend systems for different telescopes as well as between telescopes, for example any timing offsets or calibration errors since a stable system can be compared against. Such observations can thus potentially reduce the overall noise of such a combined IPTA data set.

\begin{acknowledgements}

We thank Luciano Combi and the Pulsar Monitoring in Argentina (PuMA) for useful discussions in the preparation of this work. The NANOGrav Project receives support from NSF Physics Frontiers Center award number 1430284. MTL acknowledges support from NSF AAG award number 2009468.

\end{acknowledgements}


\listofchanges

\end{document}